\documentclass[12pt]{iopart}

\expandafter\let\csname equation*\endcsname\relax
\expandafter\let\csname endequation*\endcsname\relax

\usepackage{amsmath}
\usepackage{hyperref}
\usepackage{cleveref}
\usepackage{graphicx}
\usepackage{subfig}
\usepackage{tabularx}
\usepackage{multirow}
\usepackage{booktabs}
\usepackage{amsfonts}
\usepackage{amssymb}
\usepackage{tikz}
\usetikzlibrary{calc,matrix,positioning,arrows,decorations.pathreplacing,angles,quotes,math}
\usepackage{placeins}

\begin{document}
% \mainmatter
\title[Predicting Basin Stability using Graph Neural Networks]{Predicting Basin Stability of Power Grids using Graph Neural Networks}

% \title[Predicting Dynamic Stability using Graph Neural Networks]{Predicting Dynamic Stability of Power Grids using Graph Neural Networks}
% \title[Predicting Basin Stability using Graph Neural Networks]{Predicting Basin Stability of Kuramoto-Oscillators using Graph Neural Networks}
% \title[Predicting Basin Stability using Graph Neural Networks]{Predicting Basin Stability of Swing-Equation-Oscillators using Graph Neural Networks}

\author{Christian~Nauck, Michael~Lindner, Konstantin~Schürholt, Haoming~Zhang, Paul~Schultz, Jürgen~Kurths, Ingrid~Isenhardt and Frank~Hellmann}

% \address{IOP Publishing, Temple Circus, Temple Way, Bristol BS1 6HG, UK}
% \ead{submissions@iop.org}
\vspace{10pt}
\begin{indented}
\item[]December 2021
\end{indented}

% % The paper headers
% \markboth{IEEE TRANSACTIONS ON POWER SYSTEMS}%
% {Shell \MakeLowercase{\textit{et al.}}: Bare Demo of IEEEtran.cls for IEEE Journals}

\begin{abstract}
The prediction of dynamical stability of power grids becomes more important and challenging with increasing shares of renewable energy sources due to their decentralized structure, reduced inertia and volatility. We investigate the feasibility of applying graph neural networks (GNN) to predict dynamic stability of synchronisation in complex power grids using the single-node basin stability (SNBS) as a measure. To do so, we generate two synthetic datasets for grids with 20 and 100 nodes respectively and estimate SNBS using Monte-Carlo sampling. Those datasets are used to train and evaluate the performance of eight different GNN-models. All models use the full graph without simplifications as input and predict SNBS in a nodal-regression-setup. We show that SNBS can be predicted in general and the performance significantly changes using different GNN-models. Furthermore, we observe interesting transfer capabilities of our approach: GNN-models trained on smaller grids can directly be applied on larger grids without the need of retraining.
\end{abstract}

%
% Uncomment for keywords
\vspace{2pc}
\noindent{\it Keywords}: Complex Systems, Nonlinear Dynamics, Dynamic Stability, Basin Stability, Power Grids, Machine Learning, Graph Neural Networks
%
% Uncomment for Submitted to journal title message
%\submitto{\JPA}
%
% Uncomment if a separate title page is required
%\maketitle
% 
% For two-column output uncomment the next line and choose [10pt] rather than [12pt] in the \documentclass declaration
%\ioptwocol
%
% \IEEEpeerreviewmaketitle

\section{Introduction}
The energy transition is one of the key aspects to meet the goals of the Paris Agreement \cite{united_nations_paris_2015} and its latest successor: Conference of the Parties 26 in Glasgow in 2021. Due to decentralization, reduced inertia as well as volatility in production, integrating renewable energies remains challenging. To safely operate future power grids, the impact of unavoidable fluctuations on the synchronous operating regime has to be limited. Hence, dynamic effects have to be taken into account. Analyzing the dynamic stability of synchronisation in power grids is a complex multi-dimensional problem and many known methods rely on heavy simulations. 

The model underlying the recent work on the stability of synchronization and complex dynamics of power grids, e.g. \cite{anvari_introduction_2020}, is the Kuramoto model \cite{kuramoto_self-entrainment_1975} with inertia. In complex system science it also serves as a paradigmatic model for the study of complex phenomena on networks in general \cite{acebron_kuramoto_2005, rodrigues_kuramoto_2016}. Thus, the results here are of interest beyond the specific scope of power grid modeling.

In the context of complex systems, linear stability assessments alone, e.g. based on Lyapunov exponents, are not always applicable or sufficient. A standard in the power grid community are highly detailed simulations of individual faults. For large systems, the study of all potential individual faults is too expensive, because there are too many of them. To gain a better understanding of the type of faults that might be critical, probabilistic approaches are used. They provide an appropriate understanding and heuristics for prioritizing detailed model studies to systematically investigate the dynamic stability.

Single-node basin stability (SNBS) is such a probabilistic measure. Based on the notion of the `basin of attraction' of a stable state, SNBS captures highly non-linear effects and enables the analysis of large perturbations \cite{menck_how_2013}. SNBS measures the probability of a grid to synchronize after applying sample-based-perturbations at individual nodes. SNBS has been applied to a variety of problems e.g. in the engineering community for the analysis of perturbed generators in networks \cite{liu_quantifying_2017,liu_basin_2019} and to study collective phenomena in oscillator networks \cite{rakshit_basin_2017,majhi_emergence_2019}. There are also theoretical investigations of network properties \cite{menck_how_2014,schultz_detours_2014,kim_community_2015,kim_building_2016,nitzbon_deciphering_2017,kim_multistability_2018,kim_structural_2019}. In the non-linear dynamics and complex systems community the concept of SNBS has further been analyzed and extended \cite{schultz_bounding_2018,ji_stochastic_2018,lindner_stochastic_2019,hellmann_network-induced_2020} to cover the type of dynamical properties that occur in realistic simulations of power girds, such as repeated perturbations, stochasticity and the influence of heterogeneity \cite{wolff_power_2018}. Basin stability has broadly been used to study collective phenomena in oscillator networks

Probabilistic methods like SNBS have the advantage of assessing the robustness of locations in a network independent of specific individual faults. However, as they typically rely on Monte-Carlo sampling, they are also computationally challenging. Network theoretic methods in turn already found success at predicting dynamic properties such as SNBS \cite{menck_how_2014,schultz_detours_2014,nitzbon_deciphering_2017}, raising the potential to use network theoretic heuristics to identify key structural imprints and prioritize detailed fault simulations. For example, Schultz et al. \cite{schultz_detours_2014} predict certain nodes with low SNBS using logistic regression based on network properties as input. However, the parametrization of the structure and dynamics in real power grids is highly heterogeneous, and standard network measures are not able to accommodate a wide range of node types and properties necessary for detailed, realistic dynamic models. 

Further, several network measures are not well-defined for heterogeneous systems or might not translate well from homogeneous systems.
In contrast, modern Machine Learning (ML) is able to learn complex, nonlinear patterns from any type of raw data \cite{goodfellow_deep_2016}. Hence, this work investigates the prediction of SNBS using the full graph as input.

Graph Neural Networks (GNNs) are a promising approach, because they are capable of predicting a variety of network measures \cite{avelar_multitask_2019,maurya_fast_2019} and can deal with full graphs as input. Hence, GNNs can analyze full homogeneous and heterogeneous systems without further assumptions and simplifications. Therefore, we test the prediction of SNBS using GNNs. Our paper is based on a master's thesis \cite{nauck_prediction_2020} and except of this thesis, the authors are not aware of any literature using the same methods and ideas, but we introduce related work that founds on similar approaches.

\paragraph*{Similar approaches}
There are recent publications on using Graph Neural Networks in the context of power grids, but they do not consider the prediction of statistical dynamical properties such as SNBS. Instead, many approaches deal with the computation of power flows \cite{donon_graph_2019,kim_graph_2019,bolz_power_2019,retiere_spectral_2020,wang_probabilistic_2020,owerko_optimal_2020}. GNNs have also been used for control theory \cite{gama_graph_2020} and physical neural solvers have been introduced to connect GNNs with differential equations \cite{misyris_physics-informed_2020}. Furthermore, cascading failures were investigated in \cite{liu_searching_2021}.

Aside from GNNs, two other publications are noteworthy to mention. Che et al. \cite{che_active_2021} recently published a paper in which they show the usage of active learning and relevance vector machines to reduce the computational effort of computing SNBS by learning the boundary of stable dynamics. Furthermore, Yang et al. \cite{yang_power-grid_2021} predict the ability of power grids to synchronize after applying perturbations, but they approach the concept of dynamic stability differently. Firstly, they predict the result of single perturbations and not the statistics. Secondly, their approach is not based on providing the full graph, but they rely on common knowledge about the relation of network science and dynamic stability, e.g. by using the degree and betweenness \cite{freeman_set_1977} as input. 

\paragraph*{Our main contributions}
\begin{enumerate}
    \item For the first time, SNBS is predicted based on the full graph instead of hand-crafted features. The focus lies on evaluating different learning methodologies based on GNNs for the sake of future research. The accuracy still needs to be improved for real world applications.
    \item In order to train ML-models, we generate new datasets. They are based on well-known models of synthetic power grids and on Monte-Carlo simulations to analyze dynamic stability. The datasets are rich enough to challenge ML-methods, whereas still being somewhat conceptual to connect to the existing network literature. Compared to real-world power grids, synthetic power grids have a number of advantages, for example they do not have any artifacts and one can obtain more easily large datasets, which are beneficial for statistical analyses.
    \item We also investigate transductive transfer learning capabilities by training models on small power grids and evaluating the same models on larger networks without fine-tuning.
\end{enumerate}

\paragraph*{This paper is structured as follows.} Firstly, the generation of the datasets is explained. Afterwards, the background knowledge for the used ML-methods is introduced, before we present the methodology of applying our ML-models to our generated datasets. Finally, the results are given and discussed, before we close with a short outlook.

\section{Generation of the datasets}
To analyze the capability of predicting SNBS using ML, two synthetic datasets are generated. We generate new synthetic datasets, because we are especially interested in a method that can deal with different topologies. We start by motivating the selection of our datasets. Afterwards we briefly discuss relevant concepts from network science, before explaining the generation of synthetic power grids. We close by providing details about the dynamical simulations.

\subsection{Objectives for datasets}
High-quality datasets facilitate the application of ML-methods. Therefore, we carefully consider the following criteria for generating the datasets which mimic basic features of power grids. The datasets shall be:

\begin{enumerate}
    \item homogeneous enough in both structure and dynamics to connect to network theory, 
    \item complex enough to be challenging for ML-methods,
    \item computationally feasible using highly accurate Monte-Carlo simulations.
\end{enumerate}

Firstly, homogeneity is important, because previous studies, e.g. by Nitzbon et al. \cite{nitzbon_deciphering_2017} have shown, that there are clear relations between dynamical stability and topological properties for somewhat homogeneous grids. As these patterns are known to exist in such homogeneous graph datasets, they are ideal to test ML systems, which can be expected to learn them.

Secondly, enough complexity is required to justify Machine-learning models. This complexity is inherently given in the problem setup, as SNBS is a highly non-linear measure. Furthermore, we consider different network topologies.

Thirdly, we need to find a compromise between computational effort and relevant properties of the datasets, such as grid size, number of grids, low statistical errors which are determined by the number of Monte-Carlo samples and low numerical errors, which depend on the dynamical solver settings. Low statistical errors are crucial to distinguish small performance differences between ML-models later on.

Prior to generating the datasets, the influence of many parameters is investigated. We shortly motivate and explain the most important parameters for the generation of the datasets. As previously mentioned, Nitzbon et al. \cite{nitzbon_deciphering_2017} observed interesting relations in their dataset, so we often select properties based on their investigations. Before looking at power grids in more detail, some background knowledge on graphs is needed, because power grid modeling relies on graphs.

\subsection{Network Science: graphs}
We briefly introduce theoretical background on graphs, which is also helpful to understand GNNs later on. Graphs consist of nodes (vertices) and lines (edges) connecting two nodes. The size of a graph is given by its number of nodes $N$. To encode the topology of a graph one can use the adjacency matrix $A$ which is defined by:
\begin{equation}
  A_{ij} =
  \begin{cases}
    1  & \text{if there is a line between nodes $i$ and $j$}, \\
    0 & \text{otherwise.}
  \end{cases}
  \label{eqAdjazenzmatrix}
\end{equation}
By using the degree which is defined by the number of neighbors of a node, we can formulate the diagonal degree matrix $D$. Using $A$ and $D$, we can compute the Graph Laplacian ($L$): $L = D-A$, which is a singular matrix that is a discrete analogue of the Laplace operator.

\subsection{Power grids}
The topology of the power grids is based on the tool Synthetic Networks \cite{schultz_random_2014} \footnote{This tool is available on Github \cite{schultz_luap-piksyntheticnetworks_2020}}. This package uses a parametric growth process to generate networks. The resulting networks have properties that are suitable to observations of real-world power grid networks. We use the same parametrization as Nitzbon et al. \cite{nitzbon_deciphering_2017}: $n_0 = 1, p=1/5, q=3/10, r=1/3, s=1/10$, where $n_0$ is the initial number of nodes, $p,q$ are probabilities related to constructing new lines, $s$ the probability of splitting an existing line and $r$ a parameter controlling the generation of redundant paths. Furthermore, half of the nodes are producers, whereas the other half are consumers. All nodes are modeled by the swing equation \cite{filatrella_analysis_2008}, also referred to as a second-order Kuramoto model \cite{kuramoto_self-entrainment_1975,kuramoto_self-entrainment_2005}.
The Kuramoto model was independently introduced in the context of power grids in \cite{bergen_structure_1981} and has a long history of study there. We use the following notation:
\begin{equation}
	\ddot{\phi_i} = P_i - \alpha \dot{\phi_i} - \sum_j^n K_{ij} \sin(\phi_i - \phi_j),
	\label{eqKuramoto}
\end{equation}
where $\phi,\dot{\phi}, \ddot{\phi}$ denotes the voltage angle and its time derivatives. We use the following parametrization: $P_i \in \{-1,1\}$ the injected power, whereby the condition $\sum_i P_i = 0$ guarantees power balance; $\alpha=0.1$ the damping coefficient, $K$ is the coupling matrix based on the adjacency matrix which encodes the graph and we use uniform coupling $K_{ij} = 9 A_{ij}$. The values for the injected power and the damping coefficient are the same as in \cite{nitzbon_deciphering_2017}, however we use a larger coupling (9.0 instead of 6.0) to increase the overall stability of synchronisation in the power grids and to obtain a clear bi-modal shape of the SNBS-distribution for a better balance for training ML-methods. We are interested in deviations from the nominal frequency (e.g. 50Hz in Europe), and thus will work in frequency deviations throughout the paper. The desired state is thus $\dot{\phi_i} =0$ at all nodes.

\subsection{Dataset properties}

We study the resilience of power grids operating in their synchronous state to (large) perturbations at individual nodes. The single-node basin stability of a node is quantified as the probability that the systems returns to its synchronized state after such a network-local perturbation. Since the perturbations are drawn independently at random, SNBS is the outcome of a Bernoulli experiment \cite{menck_how_2013}.

\begin{figure}
    \centering
        \subfloat{\includegraphics[width=.49\linewidth]{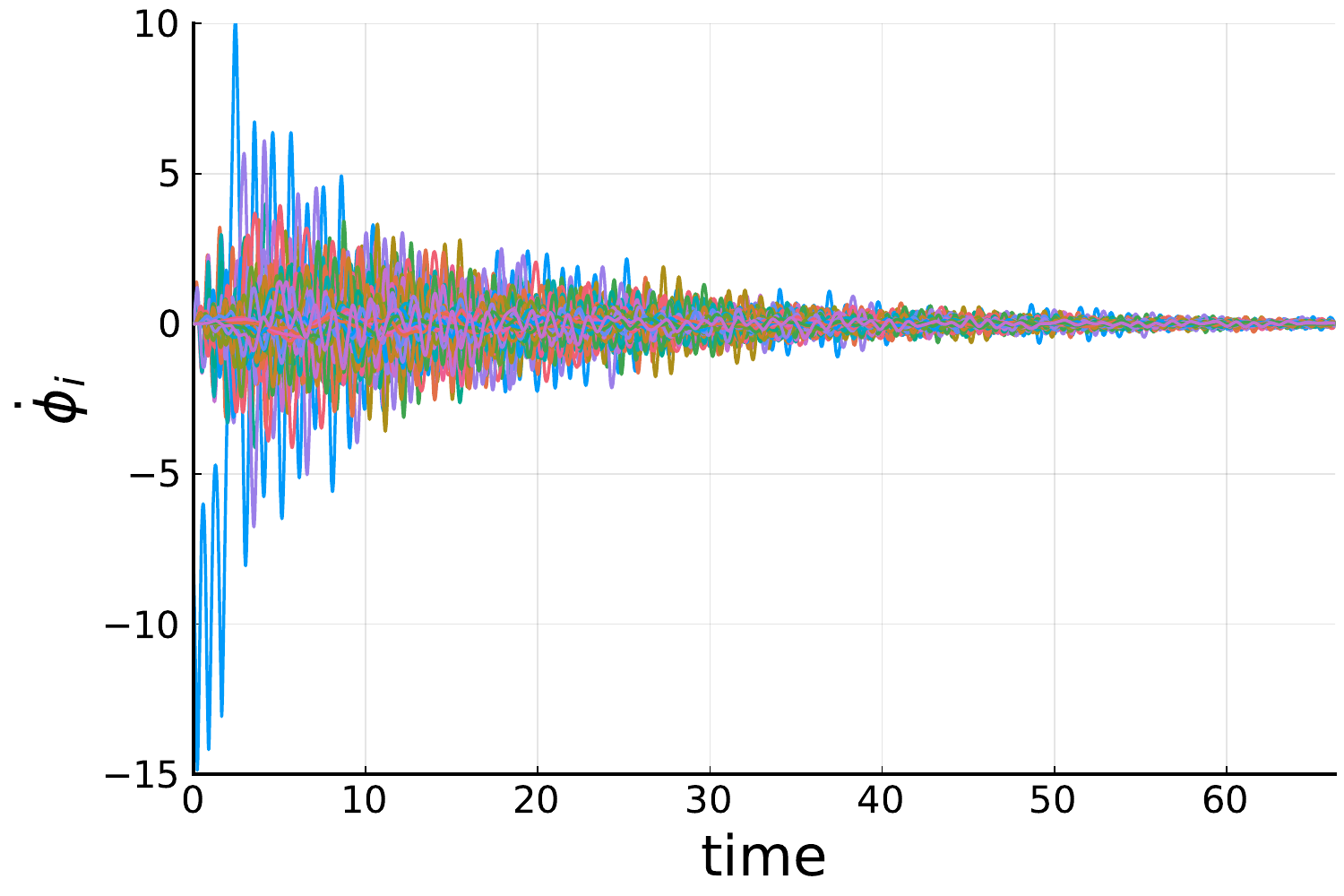}}
        \subfloat{\includegraphics[width=.49\linewidth]{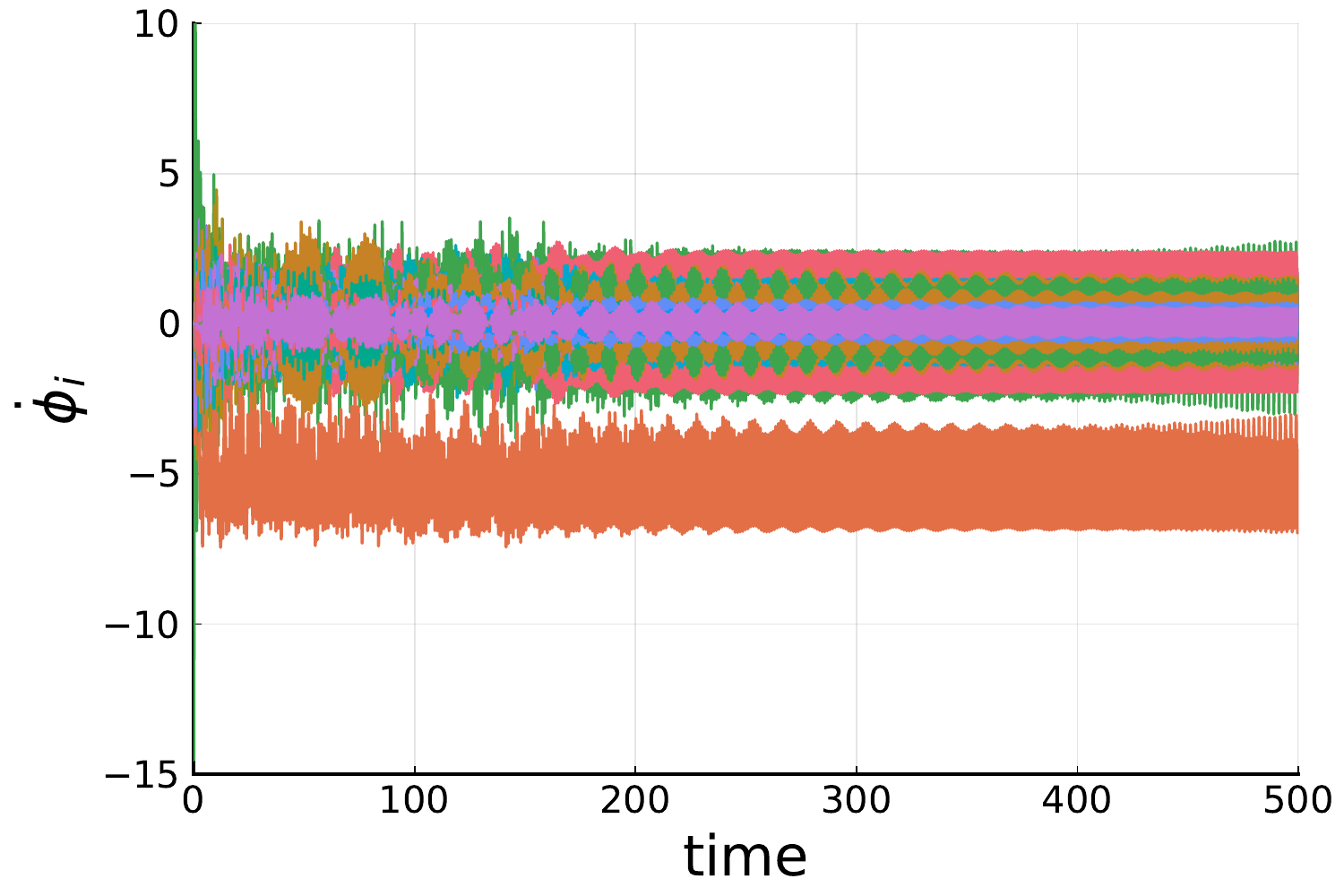}}
    \caption{Exemplary trajectories of applying single-node perturbations to a grid of 100 nodes. A stable state is reached on the left and and unstable state on the right after applying different single-node perturbations at different nodes. Different colors represent the trajectories at different nodes.}
    \label{Trajectories}
\end{figure}

To estimate SNBS for every node in a graph,  $M = 10,000$ samples of perturbations per node are constructed by sampling a phase and frequency deviation from a uniform distribution with $(\phi, \dot{\phi}) \in [- \pi, \pi] \times [-15, 15]$ and adding them to the synchronized state. Each such single-node perturbation serves as an initial condition of a dynamic simulation of our power grid model, cf. Equation~\eqref{eqKuramoto}. The simulation time is represented by $t$ in seconds. At  $t=500$ the integration is terminated and the outcome of the Bernoulli trial is derived from the final state. A simulation outcome is referred to as \emph{stable} if at all nodes $\dot{\phi_i} < 0.1$. Otherwise it is referred to as \emph{unstable}. Two exemplary trajectories are shown in \cref{Trajectories}.

The classification threshold of $0.1$ is chosen accounting for minor deviations due to numerical noise and slow convergence rates within a finite time-horizon. The authors are not aware of any other attractors of the Kuramoto system within that threshold. Hence, it may be assumed that every trajectory labeled as stable in that way will indeed converge to the synchronous state for $t \to \infty$. On the other hand, trajectories who are classified as unstable may converge to many different kinds of attractors
\cite{gelbrecht_monte_2020,halekotte_transient_2021}.
However, we occasionally observed so-called \emph{long transient states} at specific nodes, which do eventually converge to the synchronous state but fail to do so before $t=500$. While of theoretical interest, we do not expect their asymptotic behaviour to play any role in real world applications and thus we are satisfied with classifying them as unstable.

A 95\% confidence interval for the SNBS values may be estimated via the normal distribution approximation of the Bernoulli experiment as \cite{wallis_binomial_2013}:
\begin{equation}
   1.96  \sqrt{\frac{p (1-p)}{M}} < 0.01,
\end{equation}
where the inequality is obtained by setting $p=0.5$ and $M=10,000$. 

\begin{figure}
    \centering
        \subfloat{\includegraphics[width=.49\linewidth]{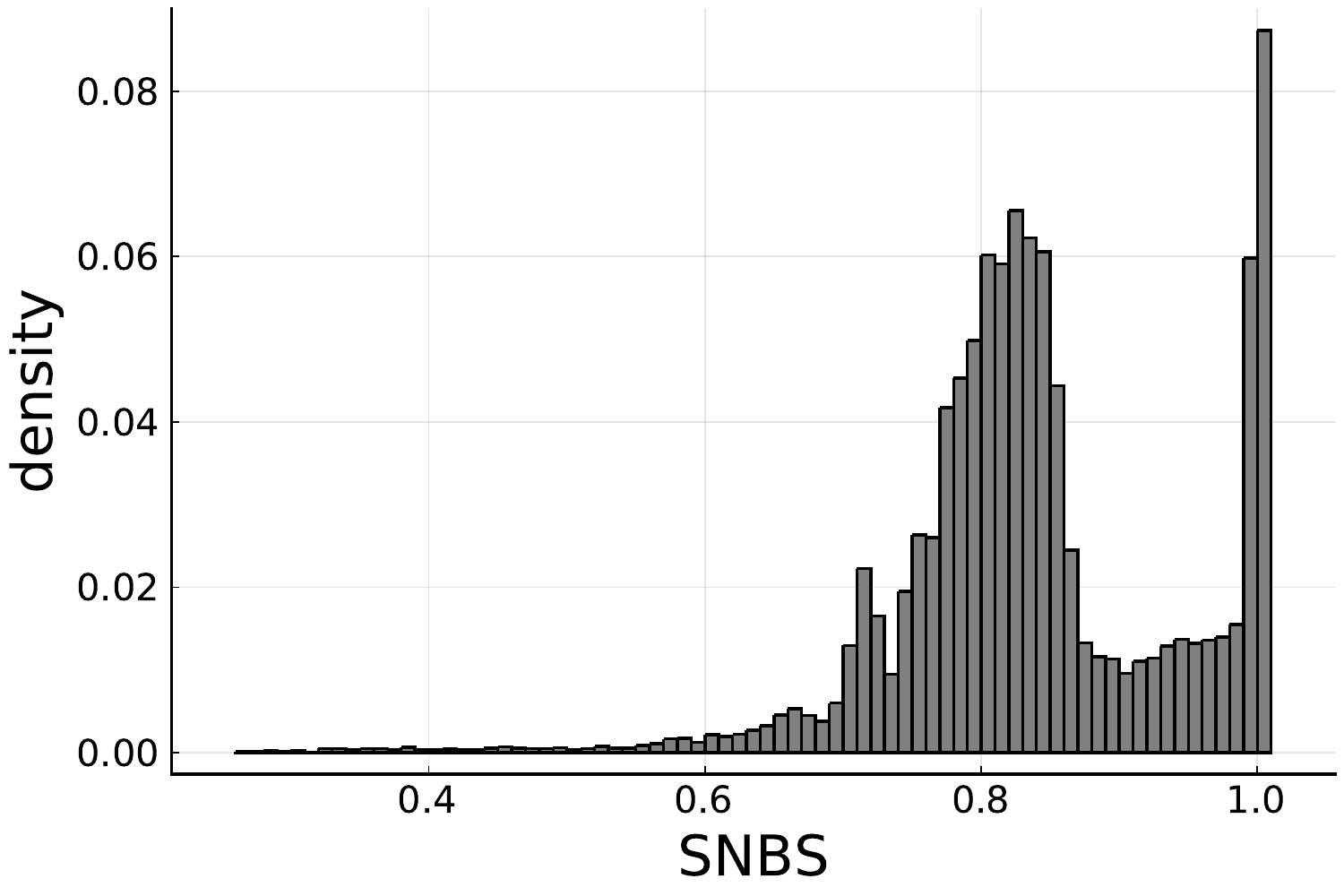}}
        \subfloat{\includegraphics[width=.49\linewidth]{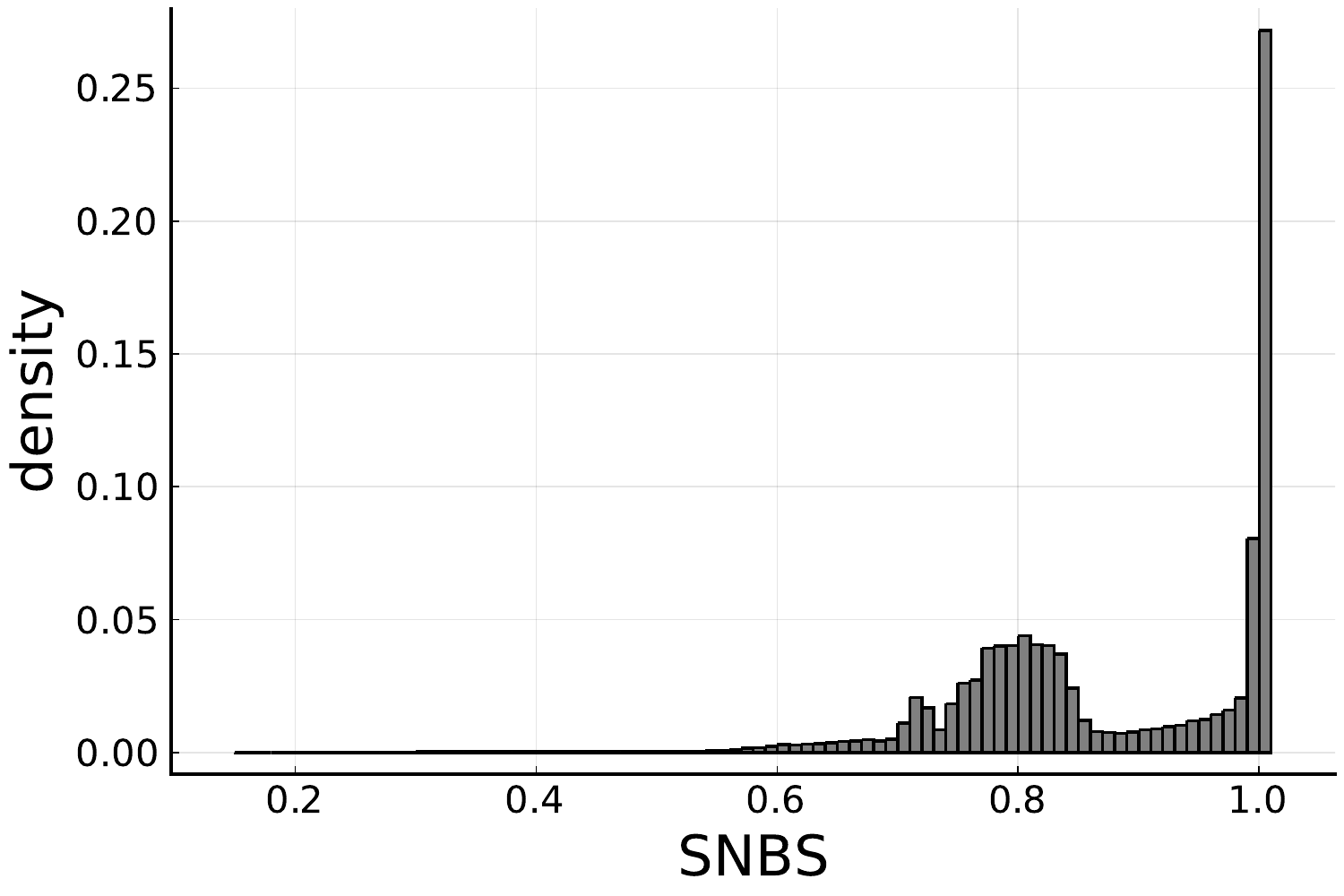}}
    \caption{Histogram showing the distribution of SNBS for the datasets with 20 nodes (left) and 100 nodes (right). The distributions are normalized so that bin heights sum to 1.}
	\label{SNBSDistribution}
\end{figure}

The distributions of SNBS for both datasets are given in \cref{SNBSDistribution}. We refer to the dataset consisting of grids of 20 nodes per grid as \emph{dataset20}, and to the dataset consisting of grids with 100 nodes as \emph{dataset100}. For both cases, there is a bi-modal distribution of SNBS over the whole data set, which facilitates ML-models to learn the distinction between those modes. The peak at $1.0$ indicates a large amount of nodes where no perturbation has an adverse effect on the synchronisation. The second peak can be interpreted in a way that many nodes are somewhat resistant to perturbations and the grid stays synchronised in about $80 \%$ when applying perturbations at the particular nodes. In case of dataset20 the mean value of SNBS is $0.84$ and for dataset100 nodes it is $0.87$. In both datasets, the number of unstable outcomes is low, which is a property we expect to hold for real power grids as well. Conducting the computation of the dynamic stability using one CPU takes about 45 hours per grid in case of 100 nodes per grid and about three hours in case of 20 nodes.

\section{Graph Neural Networks}
\label{secGNN}
This section briefly introduces Graph Neural Networks (GNNs). We begin with a general framework for GNNs and subsequently summarize the recent development of GNNs. 
Graph Neural Networks are a class of Artificial Neural Networks (ANN) designed to learn relationships of graph-structured data. Just as ANNs they have internal weights, which can be fitted in order to adapt their behavior to the given task. In the case of supervised learning these weights are adjusted such that the error between the estimated output and the labeled output for given input data is minimized. As inputs GNNs use the graph structure and potentially node features. Their output can either be global graph attributes, attributes of sub-graphs, or local node properties.
Different types of GNNs have been introduced, some of which are detailed below. In \cite{you_design_2020}, the authors introduce a \emph{design space} for GNNs as a common framework to facilitate understanding and comparison of the different methods. In their design space, GNNs consist of pre-processing, message-passing and  post-processing layers. GNN architectures vary in layer number and connectivity, as well as the intra-layer design of the message-passing layers. \cite{you_design_2020} view message-passing layers as combinations of (i) message computation and (ii) aggregation. First, a message function computes a message for each node from it's current state. Secondly, the messages are aggregated over the neighborhood to a new node state. Both message computation and aggregation can be realized in different ways. 
Common ML-methods such as \textit{batch normalization} \cite{ioffe_batch_2015} or \textit{dropout} \cite{srivastava_dropout_2014} can be added to stabilize training. The application of non-linear activation functions enables GNNs to learn non-linear relations in the graph data. 
In this work we focus on convolutional GNNs and in particular on those employing spatial-based graph convolutions, because they can be applied to varying topologies, as we have in our datasets.

Graph convolutions are based on the concept of the Graph Fourier transform, a generalization of the classical Fourier transform (FT), which enables the remarkable success of Convolutional Neural Networks (CNN) in image recognition. Unlike the classical FT, which uses exponential shifts the Graph FT corresponds to an expansion of the function on the graph in terms of the eigenvectors of the graph Laplacian $L$. Such an expansion may in turn be multiplied with a function of the graphs eigenvalues, a so-called spectral filter. While it is possible to learn spectral filters from training data, they lack many of the nice properties of the convolution kernels used in CNNs: they are not localized in node space, computing the eigenbasis is expensive and trained models can not be evaluated on different graphs, since each graph has a unique spectrum.

An important insight of \cite{hammond_wavelets_2011} was that graph spectral filters can be approximated by polynomials of the graphs' adjacency matrix $A$, thus achieving a localization of the filter in the ($k$-th order) neighborhoods of the nodes. Subsequently, in their seminal paper Kipf and Welling\cite{kipf_semi-supervised_2017} realized that it suffices to consider only the linear term of the polynomial expansion, corresponding to a simple multiplication of the node features with the (renormalized) adjacency matrix. They arrived at a computationally efficient and powerful layer architecture that relies only on local information and generalizes well to different graphs. Several GNN models that we investigate in this paper were derived from their so-called Graph Convolutional Layer (GCN):

\begin{equation}
	H = \sigma(\overline{A} X \Theta),
\end{equation}
where $H$ denotes the output of a layer, $\sigma$ is the activation function, $X$ are the input features, $\Theta$ is a matrix containing the learnable weights and $\overline{A}$ is the renormalized adjacency matrix, given by $\overline{A} = \tilde{D}^{-\frac{1}{2}} \tilde{A} \tilde{D}^{-\frac{1}{2}}$. Further $\tilde{A}=A + I$, where $I$ is the identity matrix, denotes an adjacency matrix with added self-loops and the diagonal degree matrix $\tilde{D}$ is determined by: $\tilde{D}_{ii}=\sum_j \tilde{A}_{ij}$. In the design space of \cite{you_design_2020}, $X \Theta$ manifest the message computation, while $\overline{A}$ realizes the aggregation. By consecutively applying multiple GCN-layers, not only direct neighbors are taken into account, but also neighbors at further distance. 

Instead of stacking multiple GCN-layers, Wu et al.\cite{wu_simplifying_2019} removed the activation functions, combined all weight matrices into one and computed $\tilde{A}^i$ to obtain:
\begin{equation}
	H = \textrm{softmax}(\overline{A}^i X \Theta).
\end{equation}
This layer founds on their assumption that the nonlinearity between GCN layers is not crucial and may be omitted in order to reduce computational effort. We refer to this layer as Simple-Graph-Convolution (SG).

Du et al. \cite{du_topology_2017} used multiple exponents $i$ of $\tilde{A}$ within one layer according to the following scheme:
\begin{equation}
	H = \sum_{z=0}^Z D^{-\frac{1}{2}} A^z D^{-\frac{1}{2}} X \Theta_z.
\end{equation}
This layer type is called Topology Adaptive Graph Convolution (TAG), which refers to its ability of considering different topologies. However, this is the case for all methods that are introduced in this paper. This architecture provides an extension to GCNs by incorporating information about higher order neighborhoods within one layer.

Auto-Regressive Moving Average (ARMA) neural network layers by Bianchi et al. \cite{bianchi_graph_2021} are far-reaching generalizations of GCN layers. They are derived from a rational expansion of the spectral filter instead of a polynomial expansion. A complete ARMA-layer consists itself of multiple Graph Convolutional Skip (GCS) layers:
\begin{equation}
	\overline{X}^{(j+1)}= \sigma(\tilde{L}X^{(j)}W^{(j)}+XV^{(j)}),
\end{equation}
where $j$ is an index and $W$ and $V$ are matrices of trainable parameters. There are two important distinctions from the GCN layers: the aggregation in the first term uses normalized Laplacian $\tilde{L}=I - D^{-\frac{1}{2}} A D^{-\frac{1}{2}}$, instead of $\tilde{A}$. Additionally, the connectivity of the message-passing layers is augmented with a skip connection, implemented in the second term. It recursively re-inserts the initial node features $X$ from the first layer and thus enables stacking a large number of GCS layers, whereas preventing the loss of the initial information due to Laplacian smoothing. In order to reduce the computational effort and to reduce overfitting, the weights among different GCS layers are shared: $W^{(j)}=W$ and $V^{(j)}=V$, except for the first layer where $W^{(1)} \neq W$.

To increase their expressive power multiple ARMA layers may be combined in a parallel stack: 
\begin{equation}
	\overline{X} = \frac{1}{K} \sum_{k=1}^K \overline{X}_k^{(J)},
\end{equation}
where $ \overline{X}_k^{(J)}$ is the output of the last GCS layer in the $k-$th ARMA layer. We can also interpret $J$ as the number of possible hops and by increasing $J$ larger regions are taken into account. ARMA filters with their recursive and distributed formulation, are efficient to train and capable of learning complex information. All of the layers described above are used in the models introduced in the next section.

\section{Prediction of SNBS using Graph Neural Networks}
To predict SNBS of all nodes, we use a node-regression setup, by providing the adjacency matrix of the graph and the injected power per node $P_i$ as inputs. The process is shown in \cref{ProcessPrediction}. In order to test the performance of our models on unseen data, we split the datasets into  training and testing sets. The shift between them is marginal as can be seen in \Cref{tb_datasetproperties}. 
\begin{table}[!t]
	\centering
	\caption{Properties of datasets}
	\begin{tabularx}{\linewidth}{ccccX}
		\toprule
		\multicolumn{2}{l}{name} & number of grids &number of nodes per grid & $\overline{\text{SNBS}}$ \\
		\midrule
		\multicolumn{2}{r}{dataset20} & 1000  & 20 & 0.8398\\
		& train & 800 & 20 & 0.8407\\
		& test & 200 & 20 &  0.8365\\
		\midrule
		\multicolumn{2}{r}{dataset100} & 1000  & 100 & 0.8737\\
		& train & 800 & 100 & 0.8730\\
		& test & 200 & 100 & 0.8768\\
	 \bottomrule
 	\end{tabularx}
	\label{tb_datasetproperties}
\end{table}

% \tikzstyle{input}=[circle, very thick, minimum size=0.3cm, draw=black!100, fill=black!100]
% \tikzstyle{output}=[circle, very thick, minimum size=0.3cm, draw=black!100, fill=white!100]
% \tikzstyle{layer}=[very thick, minimum size=1.5em, draw=black!100, fill=white!100, minimum
\tikzstyle{output}=[very thick, minimum size=1.5em, draw=white!100, fill=white!100, minimum width=2.5em, minimum height = 5em]
\tikzstyle{layer}=[very thick, minimum size=1.5em, draw=black!100, fill=white!100, minimum width=2.5em, minimum height = 6em]
% \begin{tikzpicture}
% 		\matrix[row sep=0.5cm,column sep=0.5cm] {
% % 		\node[label={\small $A,P$}] (inputArea) [input] {};
%         % \node[label={SNBS}] (inputArea) [output] {$\begin{bmatrix} \vdots \\ \vdots  \\ \vdots   \\ \vdots \end{bmatrix}$};
%         \node[] (inputArea) [output] {
%             % $\begin{bmatrix} \vdots \\ \vdots  \\ \vdots   \\ \vdots \end{bmatrix}$
%             \begin{minipage}{.3\linewidth}
%                 % $\begin{bmatrix} \vdots \quad \quad \vdots \\ \vdots \quad \quad \vdots  \end{bmatrix}$ \\
%                 \begin{gather*}
%                 \begin{bmatrix}
%                     \quad \quad \quad \\ 
%                      \quad A \quad \quad  \\
%                       \quad (\mathbb{R}^{N \times N})\quad \quad  \\
%                 \end{bmatrix}
%                 \\[10pt]
%                 \begin{bmatrix}
%                     \\ 
%                     P \\
%                     (\mathbb{R}^{N \times 1})\\
%                 \end{bmatrix}
%                 \end{gather*}
%             \end{minipage}
%             };
% 		&
% 		\node (layer1) [layer] {GNN};
% 		&
% 		% \node[label={\small $y$, $\bm{Y}$}] (y1) [output] {};
% % 		\node[label={SNBS}] (y1) [output] {$\begin{bmatrix} \vdots \\ \vdots  \\ \vdots   \\ \vdots \end{bmatrix}$};
% 		\node[] (y1) [output] {$\begin{bmatrix} \\ \text{SNBS}  \\ (\mathbb{R}^{N \times 1}) \\  \end{bmatrix}$};
% 		\\
% 		};
% 		\path[->]
% 			(inputArea) edge[very thick] node  {} (layer1)
% 			(layer1) edge[very thick] node  {} (y1)
% 			;
% \end{tikzpicture}

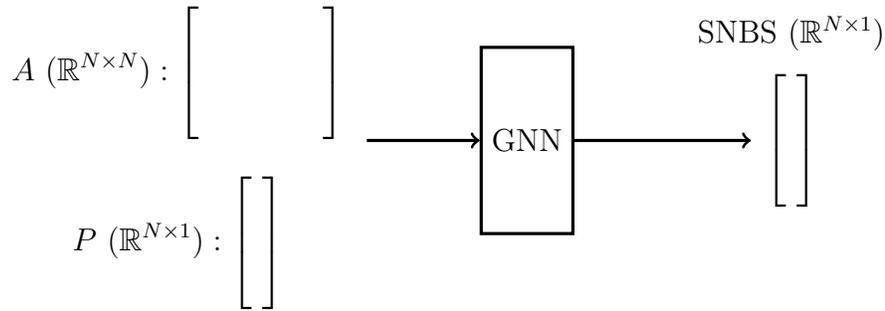
\begin{figure}
    \centering
    \begin{tikzpicture}
		\matrix[row sep=0.5cm,column sep=1.5cm] {
        \node[] (inputArea) [output] {
            \begin{minipage}{.3\linewidth}
                \begin{gather*}
                A \ (\mathbb{R}^{N \times N}): \begin{bmatrix}
                    \quad \quad \quad \quad \quad  \\ 
                     \quad  \quad \quad  \quad \quad  \\
                      \quad \quad \quad  \quad \quad  \\
                \end{bmatrix}
                \\[10pt]
                P \ (\mathbb{R}^{N \times 1}):
                \begin{bmatrix}
                    \ \\ 
                    \ \\
                    \ \
                \end{bmatrix}
                \end{gather*}
            \end{minipage}
            };
		&
		\node (layer1) [layer] {GNN};
		&
		\node[label=SNBS $(\mathbb{R}^{N \times 1})$] (y1) [output] 
		{$
        \begin{bmatrix}
            \ \\ 
            \ \\
            \ \
        \end{bmatrix}
		$
		};
		\\
		};
		\path[->]
			(inputArea) edge[very thick] node  {} (layer1)
			(layer1) edge[very thick] node  {} (y1)
			;
    \end{tikzpicture}
    \caption{Prediction of SNBS using GNN. The GNN takes the adjacency matrix $A$ and the injected power $P$ as input to obtain the nodal SNBS as output. Hence, the prediction of SNBS does not consider individual faults or any other variables, but operates only on topological properties of the power grid.}
    \label{ProcessPrediction}
\end{figure}

\subsection{Setup of our GNN-models}
Based on the introduced GNN layers, eight GNN-models are analyzed to evaluate the performance of different architectures. GNNs are capable of reading in the full graph without any simplifications. We also tried to use CNNs which are well known from image analysis. In case of CNNs, the graph information is provided by using a modified version of the adjacency matrix as input, but the setup had several limitations in comparison to the GNNs. The application of CNNs is shown in \ref{secAppCNN}. In \Cref{tb_model_properties} the GNN-models are briefly introduced. All models use one type of graph convolutional layer, but may use several numbers of them and all have one linear and one sigmoid layer at the end. Additionally, dropout is used in several cases, cf.~\ref{secModelDetails}.
We did not do a systematic investigation of hyperparameters such as number of layers and their properties, but focused on identifying relevant factors to enable training. 

\begin{table}[!t]
	\centering
	\caption{Properties of models. Number of parameters denotes the number of learnable weights of the model.}
    \begin{tabularx}{\linewidth}{XXXXX}
		\toprule
		name & type of convolution & number of layers & number of parameters & maximum number~of~hops\\
		\midrule
		ArmaNet1 & ARMA & 1 & 38 & 4\\
		ArmaNet2* & ARMA & 2 & 1050 & 8\\
		GCNNet1 & GCN & 2 & 15 & 2\\ 
		GCNNet2 & GCN & 3 & 107 & 3\\
		GCNNet3* & GCN & 3 & 149& 3\\
		SGNet1 & SG & 1 & 4 & 2\\
		SGNet2 & SG & 2 & 15 & 4\\
		TAGNet1 & TAG & 2 & 39 & 6\\
	 \bottomrule
	 \multicolumn{5}{p{\dimexpr\linewidth-2\tabcolsep-2\arrayrulewidth}}{* There is a batch normalization between first and second layer.}
 	\end{tabularx}
	\label{tb_model_properties}
\end{table}

\subsection{Training setup}
For all models the same parameters are used and the training consists of 500 epochs. To enable reproducibility, the seeds are set before training and can be found in the published source code \footnote{Information regarding the full source code is given in \ref{appSourceCode}.}. The training is based on the library Pytorch \cite{{paszke_pytorch_2019}} and for the graph handling and graph convolutional layers the additional library PyTorch Geometric \cite{fey_fast_2019} is used. For the training of the models, CPUs are used and depending on the model training takes between 20 minutes and 50 minutes on either Haswell or Broadwell architecture without parallelization. The detailed training parameters, e.g. batch sizes and additional information on the computational effort are given in \ref{secModelDetails}. As loss function we use the mean squared error \footnote{Corresponds to MSELoss in Pytorch.}.

\section{Results}
To evaluate the performance of different models, the $R^{2}$ score, which may also be known as coefficient of determination and a self-defined discretized accuracy is used. The score $R^{2}$ is computed by:
\begin{equation}
	R^2 = 1 - \frac{mse(y,t)}{mse(t_{mean},t)},
\end{equation}
where $mse$ denotes the mean squared error, $y$ the output of the model, $t$ the target value and $t_{mean}$ the mean of all considered targets of the test dataset. The standard measure of performance is $R^2$, which captures the mean square error relative to a null model that predicts the mean of the test-dataset for all points. A constant model that always predicts $t_{mean}$, disregarding the input features, would get a score of $R^2=0.0$. The $R^2$-score is used to measure the portion of explained variance in a dataset. To further simplify interpretation, we rephrase the evaluation as a classification problem. 

The outputs are categorized as true or false by using a threshold and we compute the accuracy as:
\begin{equation}
    \text{discretized accuracy} = \frac{\text{correct predictions}} {\text{number of samples}}. 
\end{equation}
We refer to this self-defined accuracy as \textit{discretized accuracy}. Predictions are considered to be correct, if the predicted output $y$ is within a certain threshold to the target value $t$: $y-t< \text{threshold}$. We set this threshold to $0.1$, because this is small enough to differentiate between the modes in the distributions (see \cref{SNBSDistribution}). Furthermore, a total deviation of the prediction and true output of $0.1$ should be efficient for most applications. The discretized accuracy depends on the distribution of SNBS, so it can not be used for comparison across different datasets, but has to be compared to the null model of the corresponding dataset.

Since there is no previous work that can be easily compared to our methods, we introduce a simple baseline model. This baseline model always predicts the average value of the testing set. By design, this results in $R^2=0$, and achieves a discretized accuracy of 67.1 \% on dataset20 and of 40.9\% on dataset100. The differences in discretized accuracy are rooted in the different distributions of the two datasets (cf. \cref{SNBSDistribution}).

\begin{figure}
    \centering
        \subfloat{\includegraphics[width=.49\linewidth]{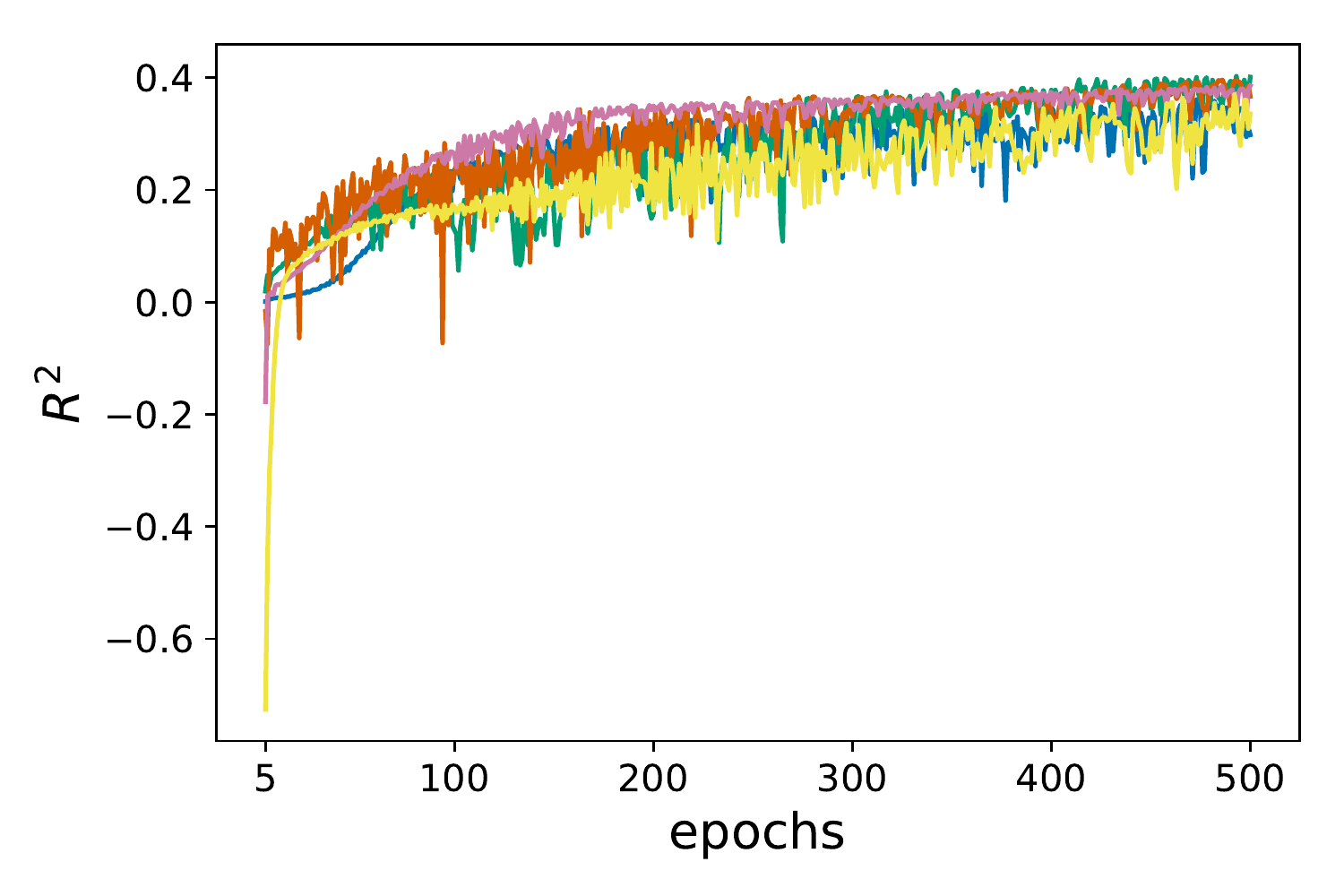}}
        \subfloat{\includegraphics[width=.49\linewidth]{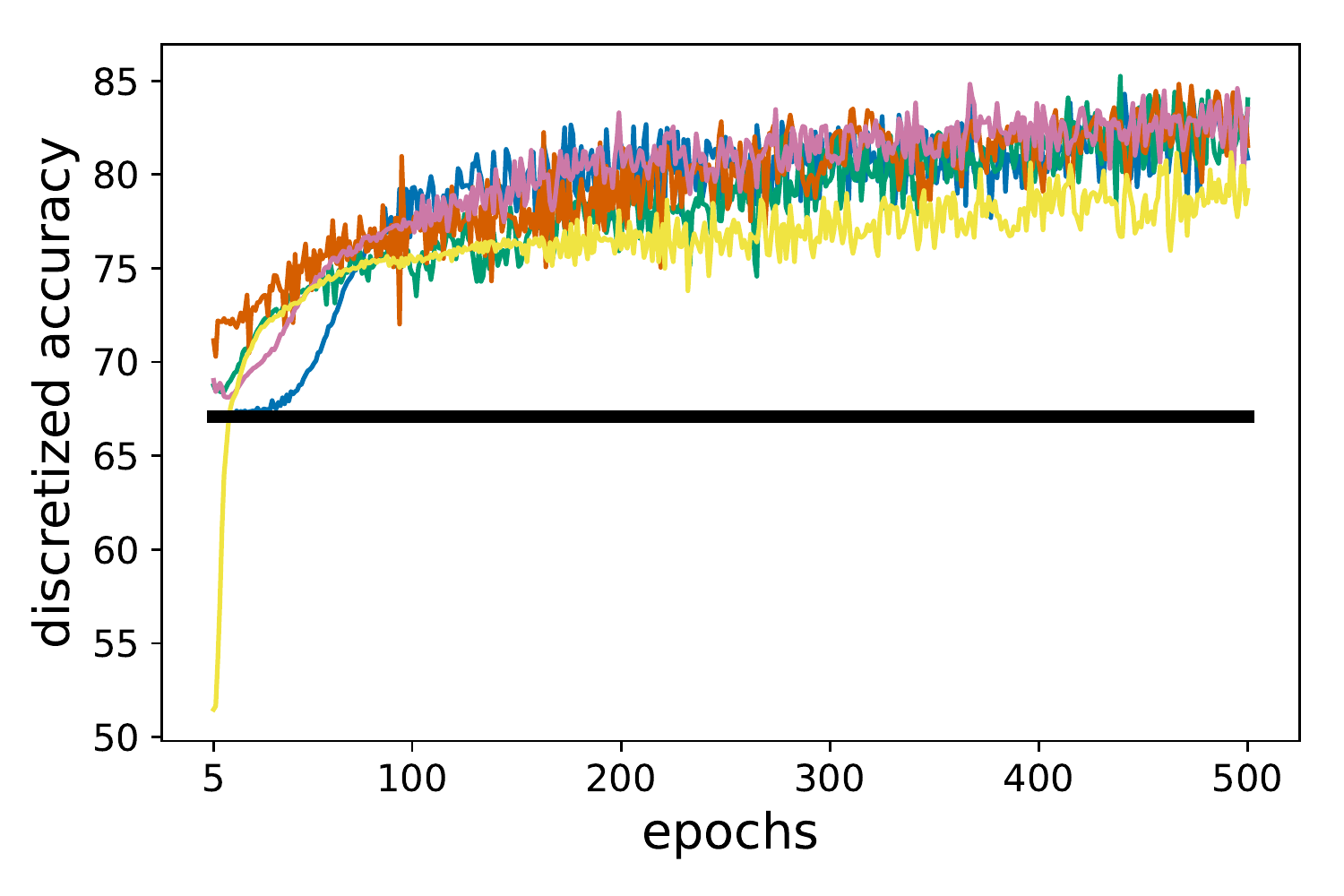}}
        \\
        \subfloat{\includegraphics[width=.49\linewidth]{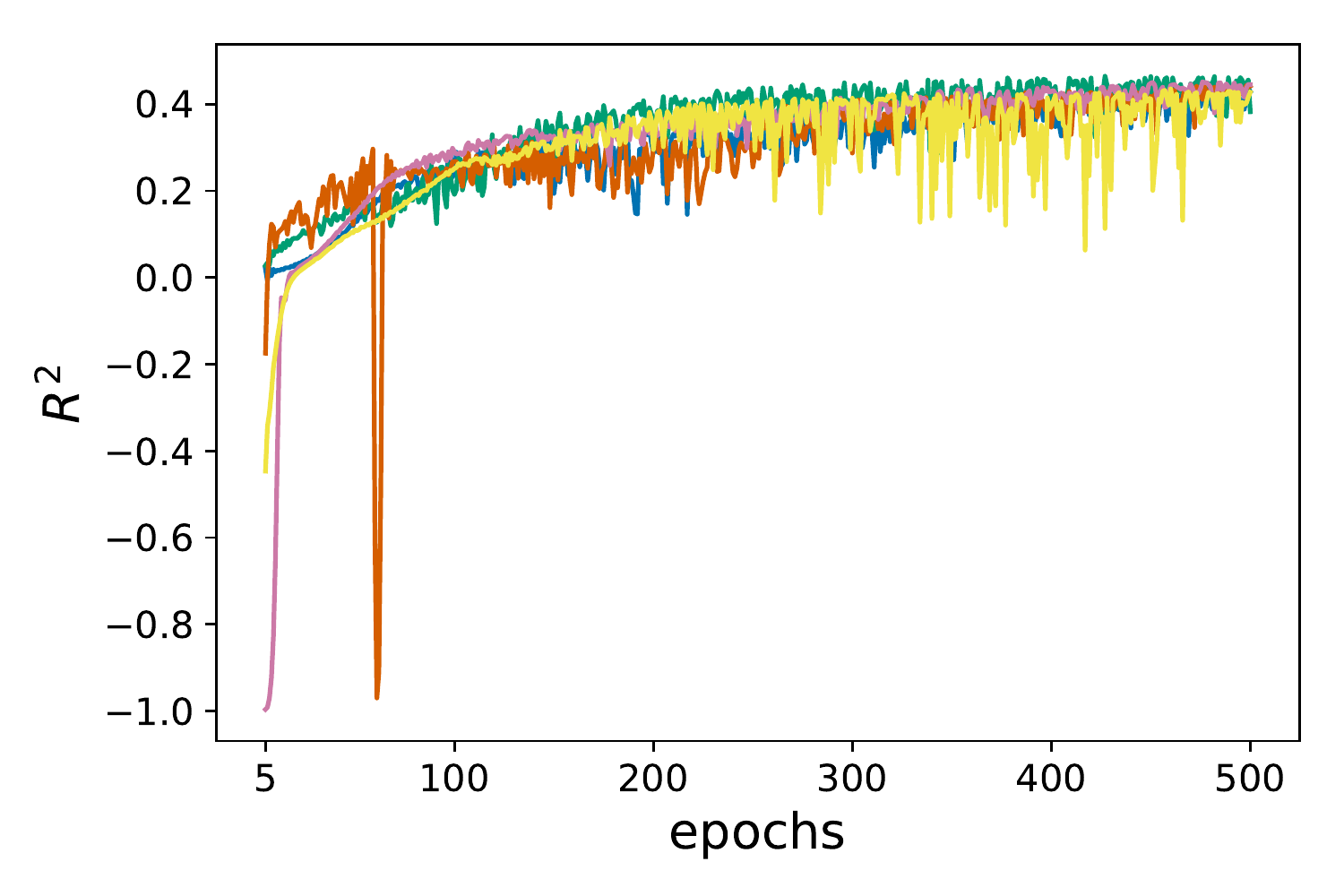}}
        \subfloat{\includegraphics[width=.49\linewidth]{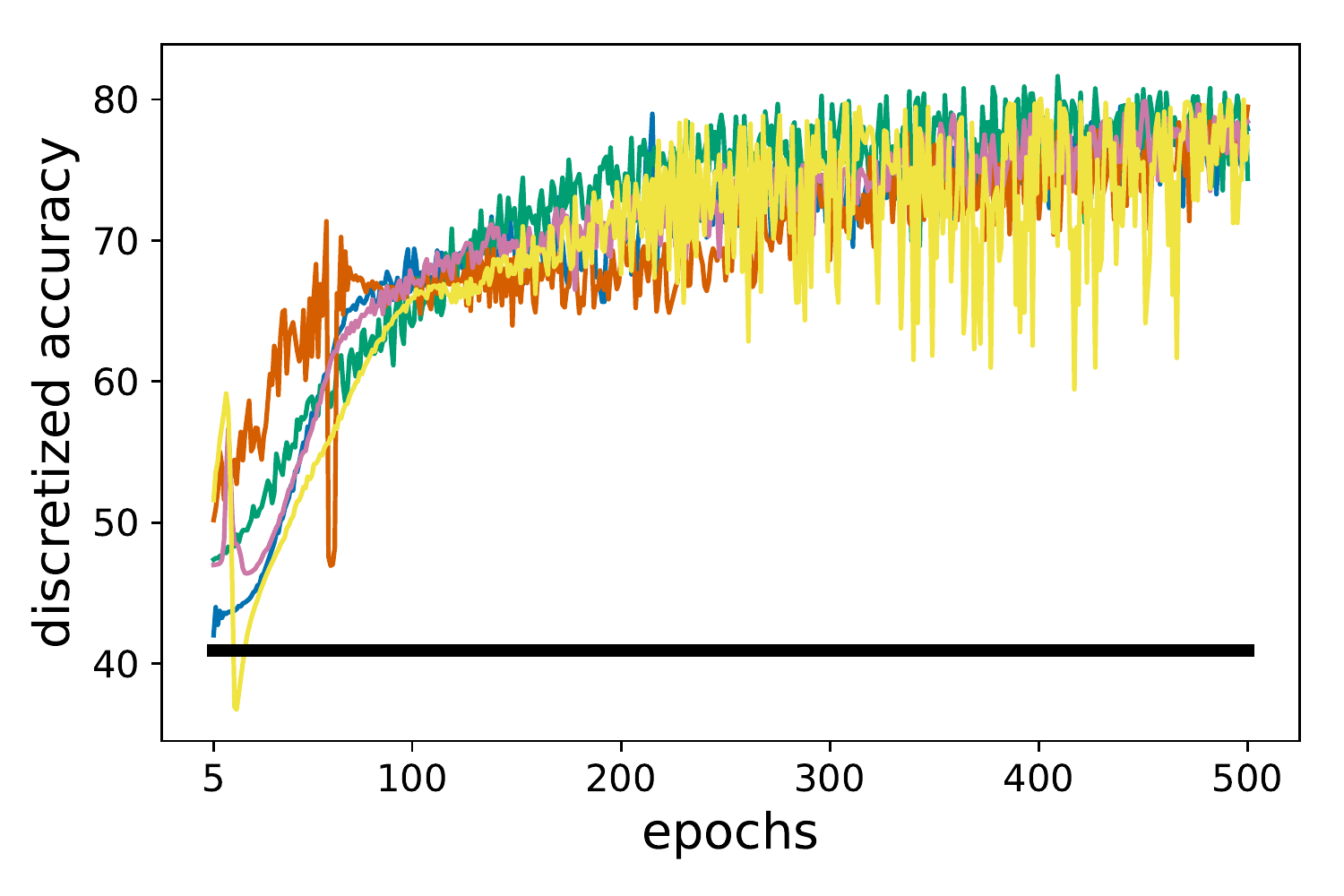}}
    \caption{Training results for ArmaNet2 and dataset20 at the top and dataset100 at the bottom. The $R^2$-score is shown on the left and the test discretized accuracy on the right. Different colors show different initializations and the horizontal line for the discretized accuracy is based on a toy model that always predicts $\overline{SNBS}$. The evaluation is purely based on the test dataset.}
 	\label{MLTrainingResultsArmaNet2}
\end{figure}

We use an averaged performance to reduce the impact of the initialization effects. Out of 5 different initializations per training setup, only the best three are considered to compute an averaged performance. The average $R^2$-performance is given in \Cref{tb_ResultsR2score} and for the discretized accuracy in \Cref{tb_ResultsAccu}. The best values are in bold. The training progress of the best model is shown in \cref{MLTrainingResultsArmaNet2}. The fluctuations, especially visible at the bottom right in \cref{MLTrainingResultsArmaNet2} are typical for ML applications when using storchastic gradient descent (SGD) and constant learning rates during training. 

Furthermore, we investigate whether the features learned by GNNs generalize to grids of different sizes. As datasets of large grids are costly to create, successful pre-training on smaller grids with subsequent application on larger grids would be a valuable strategy. To evaluate the transfer learning capabilities, we train GNN-models on the small dataset of grids with 20 nodes and evaulate without fine-tuning on the dataset with large grids of 100 nodes. As performance of the transductive transfer learning, we report the $R^2$ and accuracy on the large target dataset using the term \textit{tr20ev100} (trained on dataset20, evaluated on dataset100).

\begin{table}[!t]
	\centering
	\caption{Results represented by $R^2$ score in \%}
	\begin{tabularx}{\linewidth}{cccr}
		\toprule
	   model & dataset20 & dataset100 & tr20ev100 \\
		\hline
		ArmaNet1 & 18.8 & 15.4& 3.60  \\
		ArmaNet2 & \textbf{39.5} & \textbf{45.4} & \textbf{23.7}\\
		GCNNet1  & 8.10 & 5.98& -3.22 \\
		GCNNet2  & 24.3 & 22.1& 13.2  \\
		GCNNet3  & 9.02 & 6.71& -0.67 \\
		SGNet1   & 7.12 & 3.98& -9.15 \\
		SGNet2   & 13.5 & 13.0& 1.67  \\
		TAGNet1  & 29.1 & 28.8& 13.7  \\
	 \bottomrule
	 \multicolumn{4}{p{\dimexpr\linewidth-2\tabcolsep-2\arrayrulewidth}}{For dataset20 and dataset100, the models are both trained on their training and evaluated on their test sections. To evaluate the transfer learning capabilities, we use the term \textit{tr20ev100} meaning that the model is trained on the dataset20, but evaluated on the dataset100.}
 	\end{tabularx}
 	\label{tb_ResultsR2score}
\end{table}

\begin{table}[!t]
	\centering
	\caption{Results represented by discretized accuracy in \%}
	\begin{tabularx}{\linewidth}{cccr}
		\toprule
	   model & dataset20 & dataset100 & tr20ev100\\
		\hline
		ArmaNet1 & 76.5 & 65.1& 56.0\\
		ArmaNet2 & \textbf{80.5} & \textbf{85.0} & \textbf{65.9}\\
		GCNNet1  &  69.5 & 66.6& 47.8\\
		GCNNet2  &  79.8 & 67.5& 59.8\\
		GCNNet3  &  71.6 & 63.7& 49.5\\
		SGNet1   &  67.9 & 67.8& 46.0\\
		SGNet2   &  70.5 & 65.9& 48.7\\
		TAGNet1  &  78.8 & 69.6& 56.1\\
	 \bottomrule
	 \multicolumn{4}{p{\dimexpr\linewidth-2\tabcolsep-2\arrayrulewidth}}{For dataset20 and dataset100, the models are both trained on their training and evaluated on their test sections. To evaluate the transfer learning capabilites, we use the term \textit{tr20ev100} meaning that the model is trained on the dataset20, but evaluated on the dataset100.}
 	\end{tabularx}
	\label{tb_ResultsAccu}
\end{table}

\begin{figure}
    \centering
        \subfloat{\includegraphics[width=.49\linewidth]{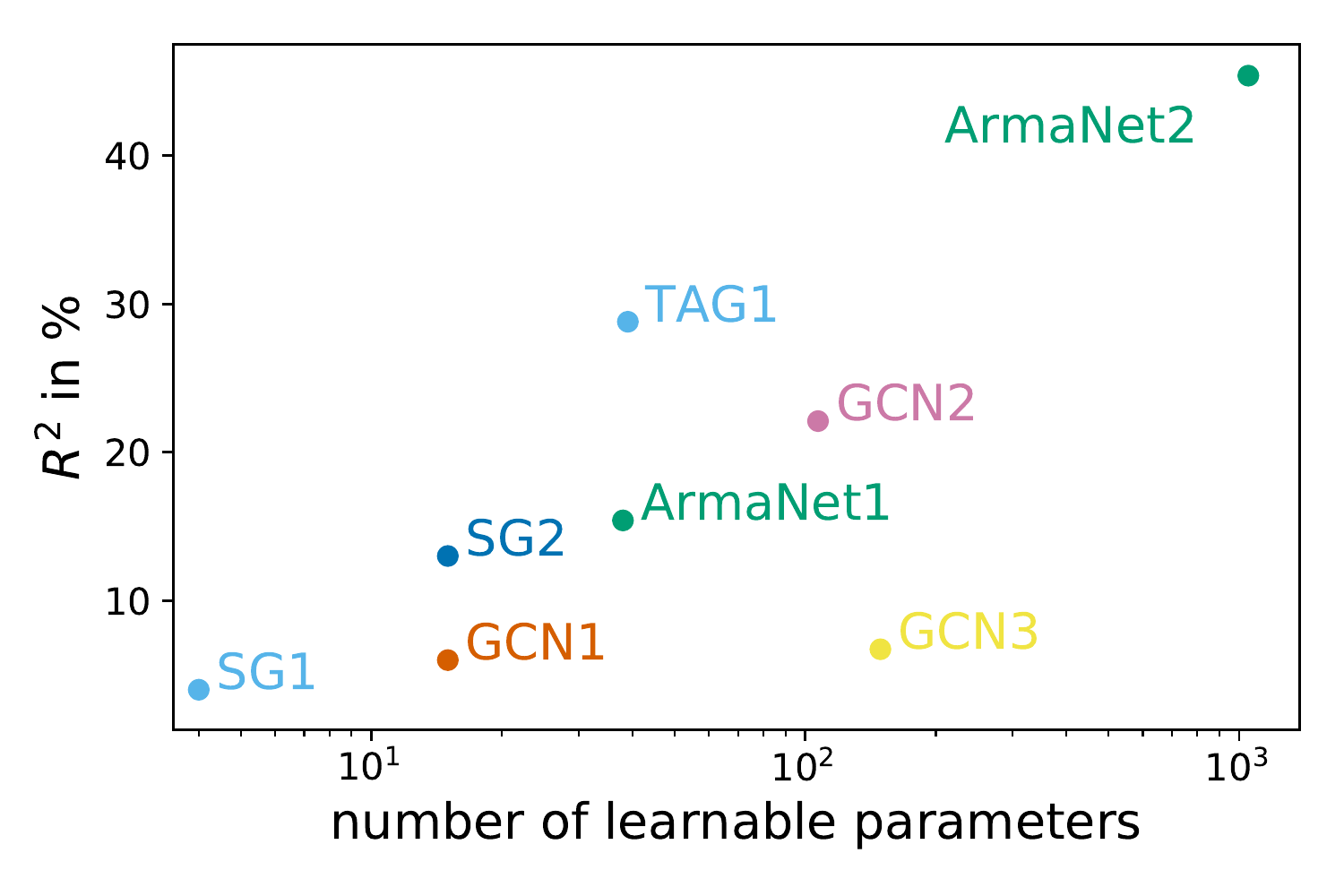}}
        \subfloat{\includegraphics[width=.49\linewidth]{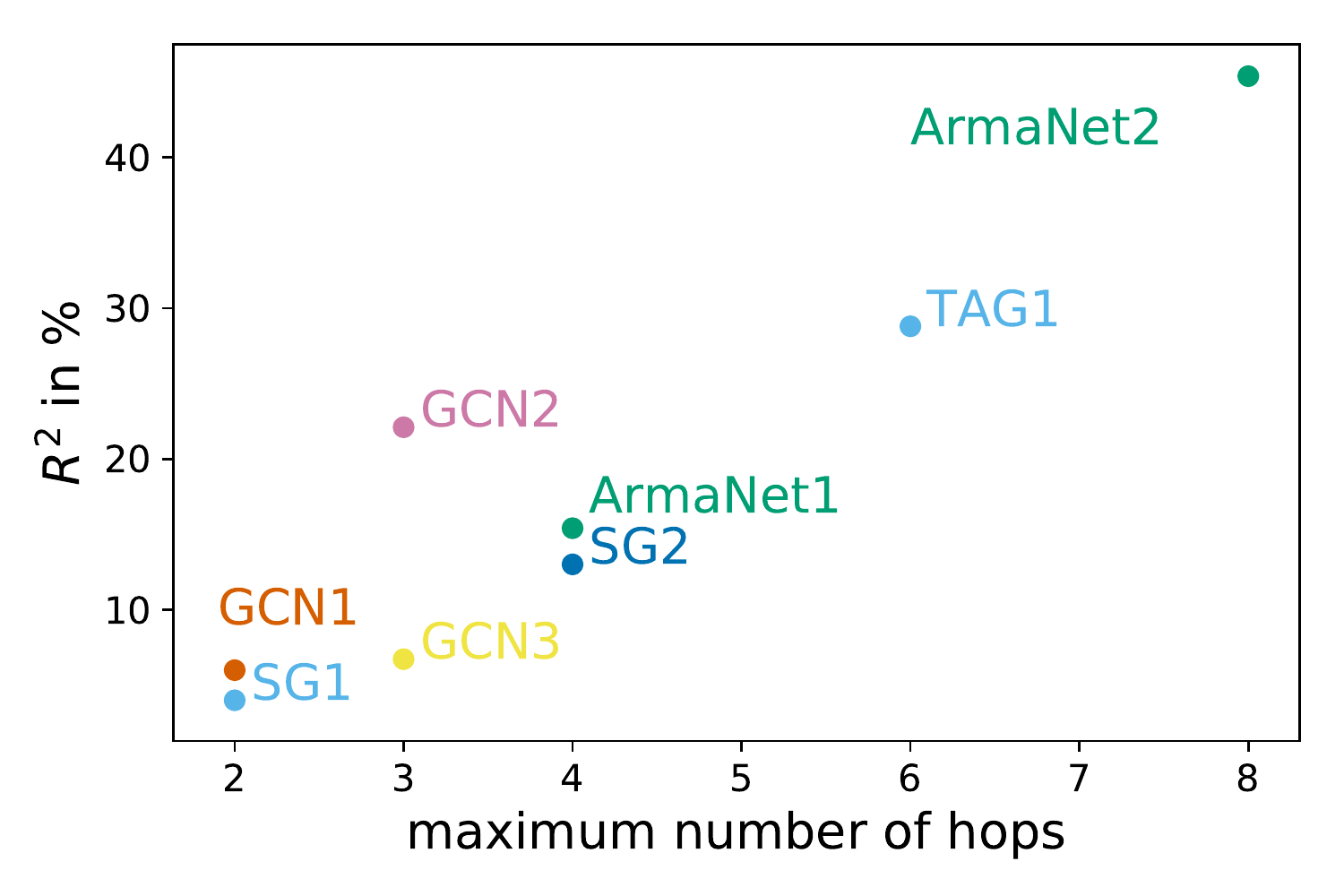}}
    \caption{Relation of performance and the complexity of models represented by the number of learnable parameters on the left and the number of maximum hops on the right. The plotted results are based on dataset100.}
	\label{complexity_vs_performance}
\end{figure}

The results show that the prediction of SNBS using GNNs is feasible and different models have a large impact. We did not perform a detailed hyperparameter study of different GNN-models, so conclusions about their performance are tentative for now. Next, we shortly summarize our observations. The results indicate that increasing the complexity of the model can be beneficial, as the model ArmaNet2 with the largest amount of parameters (1050) performs best. However, increasing the complexity is not always helpful. GCNNet3 for example performs worse than GCN2, even though having more learnable parameters (149 instead of 107). The meaning of the type of convolution is underlined by considering TAGNet1 and ArmaNet1, because TAGNet1 outperforms ArmaNet1 with only slightly more parameters than ArmaNet1. \Cref{complexity_vs_performance} shows the relation of the complexity and performance based on dataset100. The complexity is firstly represented by the number of learnable parameters on a logarithmic scale and secondly by the maximum number of possible hops. By hops we mean the order of neighbors that are taken into account. For example, one hop means that only direct neighbors are considered, whereas two means that nodes are considered which are not directly connected, but via one direct neighbor. 

Without conducting ablation studies, we can only guess reasons for the superiority of ArmaNet2. We suspect two main reasons: Firstly, the largest number of parameters could be decisive; Secondly, the most complex architecture including skip layers to consider neighbors of higher degrees could have a positive impact. The four GCS-layers of ArmaNet2 can consider a relatively large region. TAGNet1 also performs well and this model can evaluate neighbors of $6^{th}$-order, by having two layers and three hops per layer. The benefit of ArmaNet2 can be emphasized by investigating tr20ev100, because ArmaNet2 outperforms all other models on dataset100, even if it is purely trained on dataset20. Hence, the models ArmaNet2 results in the most robust setup.

\begin{figure}
    \centering
        \includegraphics[width=.98\linewidth]{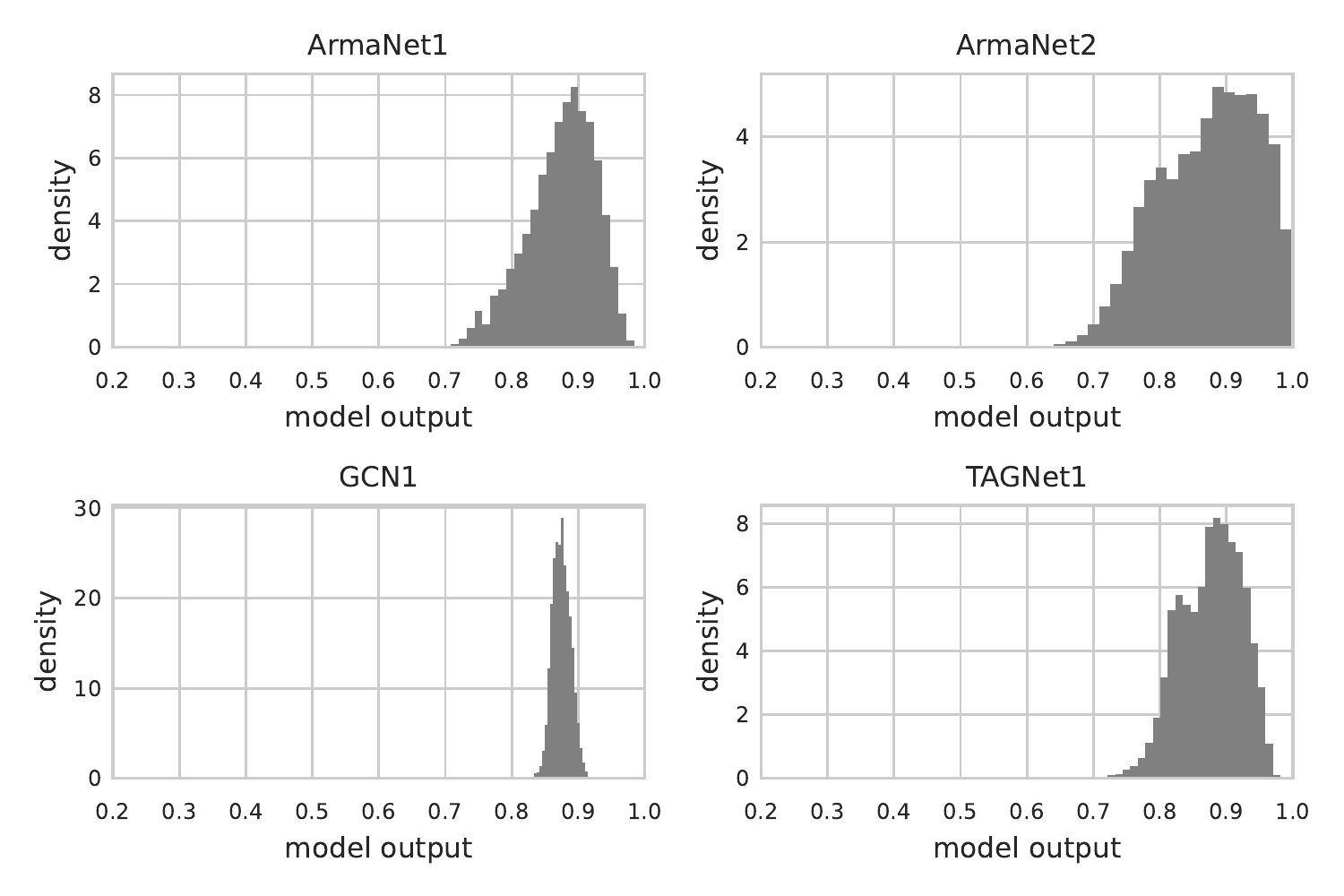}
    \caption{Histograms showing density of predicted outputs for different models and dataset100 and the best seed per model.}
    \label{ModelComparison_Histogram}
\end{figure}

To further evaluate the performance of the investigated models, we analyze the distribution of the output of selected models in \cref{ModelComparison_Histogram}. Therefor, we only consider the output based on the best seed per model using $R^2$ as a criterion. The output of all models is restricted to somewhat large values and neither low nor very high values of SNBS can be predicted. The small amount of nodes with low SNBS in the dataset might explain the absence of low output values. In case of large output values, it is remarkable and a bit surprising that none of the models predicts the abundance of completely stable outcomes. This behaviour limits the applicability to real world problems. The limitation of all models also becomes clear when comparing the results to the distributions introduced in \cref{SNBSDistribution}. Since the shifts\footnote{A dataset shift means, that training and testing datasets are different.} within the datsets are small, we can compare the output distributions to the distributions of the entire datasets, even though \cref{ModelComparison_Histogram} only considers the test section.

The distributions of the output (\cref{ModelComparison_Histogram}) also indicate performance differences between the models. We clearly see that GCN1, having a relatively low performance, has a very limited range of output values and all values are around the mean of the dataset. ArmaNet1 already has a wider range, whereas ArmaNet2 has the largest range. Besides the range, the shape of the distribution and modalities of the predictions are also telling, e.g. we find an indication for a bimodul distribution in case of TAGNet1. All in all, the superiority of ArmaNet2 and TAGNet1 becomes visible. However, even for those models the output is still limited to values that are larger than 0.6 and there is only a small amount of predictions of high stability (SNBS$\approx $1).

\begin{figure}
    \centering
        \includegraphics[width=.98\linewidth]{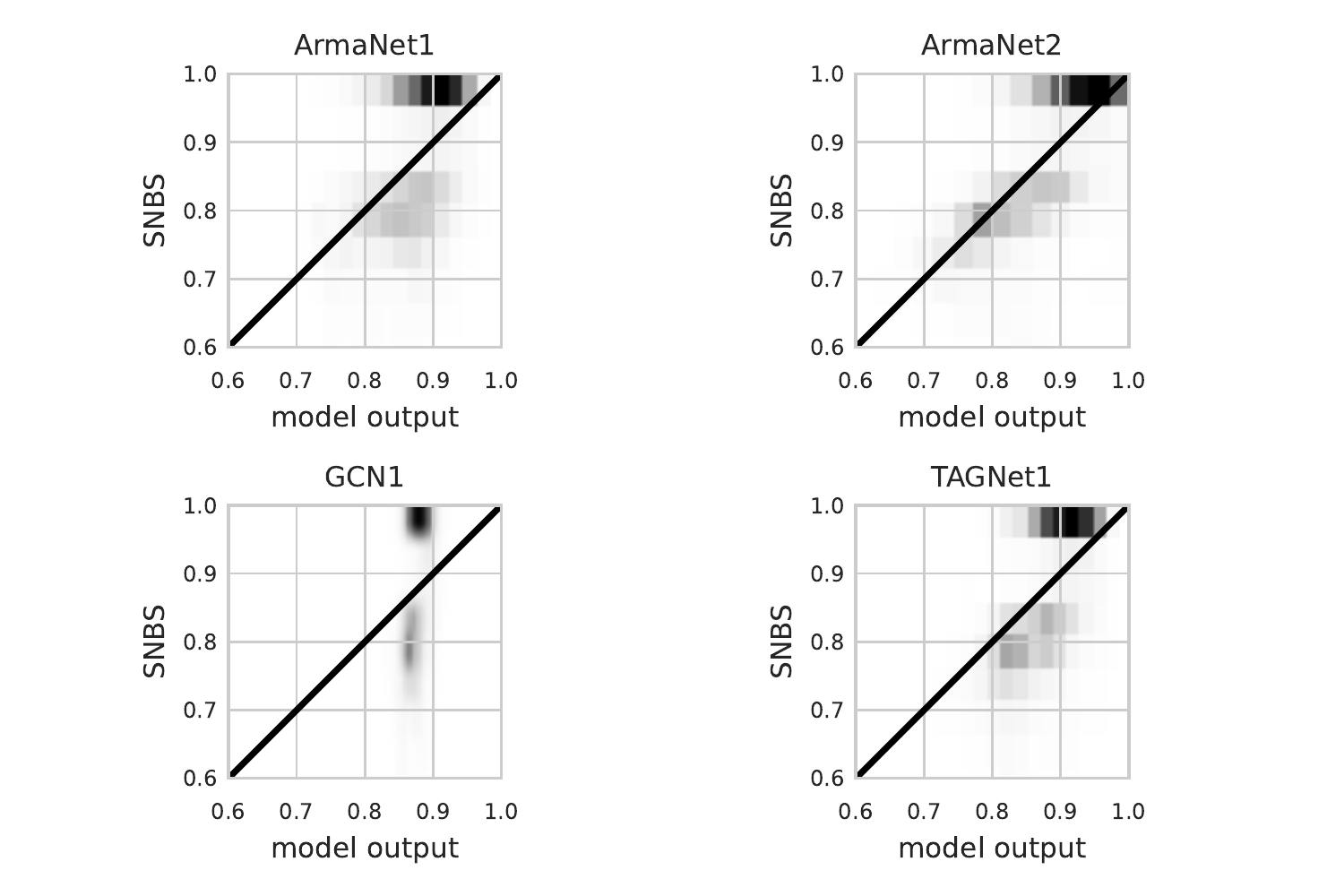}
    \caption{Heat maps of comparing models using the best seed for each of them and considering the predicted output vs. SNBS and investigating dataset100. The diagonal represents a potential perfect model ($R^2=1$).}
    \label{ModelComparison_Output_vs_labels_heatmap}
\end{figure}

To visually analyze the models, we plot the predicted output vs. SNBS in heat maps in \cref{ModelComparison_Output_vs_labels_heatmap}. Perfect predictions would be on the diagonal only, similarly to $R^2=1$. On the contrary to $R^2$ shown in \cref{tb_ResultsR2score}, we can find some reasons for the performance differences. We see that ArmaNet2 and TAGNet1 can distinguish between nodes with SNBS $\approx 1$ and nodes with lower SNBS. Other models, such as GCN1, have large regions on the off-diagonal, resulting in a lower performance.


\section{Conclusion and Outlook}

The key result of this paper is a novel approach of estimating SNBS via GNNs. We have demonstrated its potentials and have paved the way for further investigations. We show the necessity to use well-adapted architectures for this problem, since generic CNNs are not able to achieve comparable results even with more parameters (cf.~\ref{secAppCNN}).

The strongest limitation of the presented results are probably the assumptions for generating the datasets which matches several properties of real power grids, but it also simplifies some aspects, e.g. missing heterogeneity of nodes (power input) and lines (coupling constant). However, the accuracy can still be increased before moving to more realistic setups, because the performance is still too low for real applications. We provide several ideas for improvements in the next paragraphs.

Since we see substantially improved performance for models with larger number of parameters testing more complex models seems very promising. More complex models might identify other relevant structures of networks to predict SNBS more accurately, there is no suggestion that the performance is already saturating. As a first step, one could conduct a hyperparameter study to improve the investigated models.

In further steps, one could introduce new models to increase the performance. Firstly, new layers could be designed that specifically aim to predict SNBS and deal with power grids. Secondly, hybrid approaches might be used that incorporate knowledge about known structures, e.g. network motifs that can hardly be recognized by GNNs. Generally it is clear from our results that more complex architectures are promising for this task, even if it remains unclear exactly what direction the complexity increase should point towards.

Another key for improvement are the datasets. The used datasets are relatively small, so increasing the size of the datasets might be an important step for training more complex models. To solve the issue of the limited range of outputs and the observation that the model outputs are around the mean of the datasets, balancing or weighting of samples might help. 

Remarkably, we successfully showed that GNNs can generalize across different sizes of power grids. Another avenue for future research is to train models based on different sizes to start with. It is feasible that the overall performance can be increased when actually training the models on multiple datasets. The capability of training models on smaller grids and applying them on larger grids can become crucial for real-world applications to reduce the computational effort of generating datasets and also of training the models.

\section*{Acknowledgments}
All authors gratefully acknowledge the European Regional Development Fund (ERDF), the German Federal Ministry of Education and Research, and the Land Brandenburg for supporting this project by providing resources on the high-performance computer system at the Potsdam Institute for Climate Impact Research. The authors also thank the Chair of Information Management in Mechanical Engineering of RWTH Aachen University for computational resources. Christian Nauck would like to thank the German Federal Environmental Foundation (DBU) for funding his PhD scholarship and Professor Raisch from Technical University Berlin for supervising his PhD. Michael Lindner greatly acknowledges support by the Berlin International Graduate School in Model and Simulation based Research (BIMoS) of TU Berlin. This work was funded by the Deutsche Forschungsgemeinschaft (DFG, German Research Foundation) – KU 837/39-1 / RA 516/13-1. The publication was supported by the DFG funding program Open Access Publication Funding.

\FloatBarrier

% \ifCLASSOPTIONcaptionsoff
%   \newpage
% \fi

\section*{References}
\bibliographystyle{IEEEtran}
% \bibliography{./Literature.bib}
% !!! change to input *.bbl
% Generated by IEEEtran.bst, version: 1.14 (2015/08/26)

\newpage
\appendix

% \appendices
\section{Source code}
\label{appSourceCode}
The full source code including the dataset is available at \url{https://zenodo.org/record/5148085}. Furthermore, the scripts are also given at Github \url{https://github.com/PIK-ICoNe/paper-companion_predicting-snbs-using-gnn}. The code for the computation of SNBS is written in Julia \cite{bezanson_julia_2017} and the dynamic simulations rely on the package DifferentialEquations.jl \cite{rackauckas_differentialequationsjl_2017}. For simulating more realistic power grids in future work we recommend the additional use of NetworkDynamics.jl \cite{lindner_networkdynamicsjlcomposing_2021} and PowerDynamics.jl \cite{plietzsch_powerdynamicsjl_2021}. Software packages used for ML-applications are listed in \Cref{tb_Software}.

\begin{table}[!b]
	\centering
	\caption{Software packages for ML}
	\begin{tabularx}{\linewidth}{XXXc}
		\toprule
        package                 & version  &    package                 & version  \\
        \hline
        Cuda                    & 10.2     &    torch-cluster           & 1.5.9    \\
		h5py                    & 2.10.0   &    torch-geometric         & 1.7.0    \\
		numpy                   & 1.19.2   &    torch-scatter           & 2.0.6    \\
        pandas                  & 1.2.4    &    torch-sparse            & 0.6.9    \\
        python                  & 3.8.5    &    torch-spline-conv       & 1.2.1    \\
        pytorch                 & 1.7.1    &    torchvision             & 0.8.2    \\
	 \bottomrule
 	\end{tabularx}
	\label{tb_Software}
\end{table}

\section{Model and training details}
\label{secModelDetails}
In this section details about the used models are provided, whereby NIC denotes the number of input channels, NOC the number of output channels, NOH the number of hops per layer (for TAGConv and SGConv) and ReLU the rectified linear unit activation function. Recall that we only have one node feature as input and hence our most common choice for NIC is 1.

\small
\begin{enumerate}
\item \emph{ArmaNet1}:
\begin{itemize}
	\item Arma-Convolution (NIC = 1 , NOC = 1, num-stacks = 3, num-layers = 4, ReLU activation, weights are not shared between layers),
	\item fully connected layer and sigmoid output layer.
\end{itemize}
\item \emph{ArmaNet2}:
\begin{itemize}
	\item Arma-Convolution (NIC = 1, NOC = 16, num-stacks = 3, num-layers = 4, dropout = 0.25, ReLU activation, weights are shared between layers),
	\item Batch normalization, ReLU and dropout,
	\item Arma-Convolution (NIC = 16, NOC = 1, num-stacks = 3, num-layers = 4, dropout = 0.25, no activation function, weights are shared between layers),
	\item fully connected layer and sigmoid output layer.
\end{itemize}

\item \emph{GCNNet1}:
\begin{itemize}
	\item GCN-convolution (NIC = 1, NOC = 4),
	\item ReLU and dropout,
	\item GCN-convolution (NIC = 4, NOC = 1),
	\item fully connected layer and sigmoid output layer.
\end{itemize}

\item \emph{GCNNet2}:
\begin{itemize}
	\item GCN-convolution (NIC = 1, NOC = 16),
	\item ReLU and dropout,
	\item GCN-convolution (NIC = 16, NOC = 4),
	\item ReLU,
	\item GCN-convolution (NIC = 4, NOC = 1),
	\item fully connected layer and sigmoid output layer.
\end{itemize}

\item \emph{GCNNet3}:
\begin{itemize}
	\item GCN-convolution (NIC = 1, NOC = 16),
	\item batch normalization, ReLU and dropout,
	\item GCN-convolution (NIC = 16, NOC = 4),
	\item batch normalization, ReLU
	\item GCN-convolution (NIC = 4, NOC = 1),
	\item fully connected layer and sigmoid output layer.
\end{itemize}

\item \emph{SGNet1} :
\begin{itemize}
	\item SG-Convolution (NIC = 1, NOC = 1, NOH = 2),
	\item ReLU, fully connected layer and sigmoid output layer.
\end{itemize}

\item \emph{SGNet2}:
\begin{itemize}
	\item SG-Convolution (NIC = 1, NOC = 4, NOH = 2),
	\item ReLU, dropout,
	\item SG-Convolution (NIC = 4, NOC = 1, NOH = 2),
	\item fully connected layer and sigmoid output layer.
\end{itemize}

\item \emph{SGNet3}:
\begin{itemize}
	\item SG-Convolution (NIC = 1, NOC = 16, NOH = 2),
	\item ReLU, dropout,
	\item SG-Convolution (NIC = 16, NOC = 4, NOH = 2),
	\item ReLu,
	\item SG-Convolution (NIC = 4, NOC = 1, NOH = 2),
	\item fully connected layer and sigmoid output layer.
\end{itemize}

\item \emph{TAGNet1}:
\begin{itemize}
	\item TAG-Convolution (NIC = 1, NOC = 4, NOH = 3),
	\item ReLU, dropout,
	\item TAG-Convolution (NIC = 4, NOC = 1, NOH = 3),
	\item fully connected layer and sigmoid output layer.
\end{itemize}
\end{enumerate}

\normalsize

The used training parameters can be found in \Cref{tb_TrainingParameters}.
The scripts include seeds for \textit{torch, cuda and numpy.random}, even though cuda may not be used. % For dataset20 the training takes between 20 and 25 minutes using one CPU and for dataset100 it takes between 28 and 47 minutes.  %The computation times are given in \Cref{tb_gnn_training_effort}. The times should only be a rough estimation. 

\begin{table}[!b]
	\centering
	\caption{Parameters for ML}
	\begin{tabularx}{\linewidth}{ccXc}
		\toprule
	   parameter & property & parameter & property \\
		 \hline
		 training batchsize & 100 & test batchsize & 200 \\
		 trainig set index & 1-800 & test set index & 801-1000 \\
		 train set shuffle & true & test set shuffle & false \\
		 optimizer & stochastic gradient descent (SGD) & learning rate & 0.3 \\
		 momentum & 0.9 & weight decay & 1e-9 \\
		 criterion & MSELoss & threshold for discretized accuracy & 0.1 \\
	 \bottomrule
 	\end{tabularx}
	\label{tb_TrainingParameters}
\end{table}

\section {Convolutional Neural Networks}
\label{secAppCNN}
We used CNNs taking the graph Laplacian and information about the node type (source/sink) as input. The power is either added in an additional row to $L$, resulting in a 1-input-channel setup (1C) or concatenated as a second input channel (2C).

The results are given in \cref{tb_ResultsR2scoreCNN}. The input formats (1C and 2C) do not have a large impact. All CNNs achieve comparable performance and outperform low performing GNN-models, but they are not competitive to the best GNN-models.

\begin{table}[!b]
    \caption{Results when using CNNs }
	\centering
	\begin{tabularx}{\linewidth}{Xcccccccc}
		\toprule
		model & \multicolumn{4}{c}{$R^2$ score in \%} &\multicolumn{4}{c}{discretized accuracy in \%}\\
	   & \multicolumn{2}{c}{dataset20} & \multicolumn{2}{c}{dataset100} &\multicolumn{2}{c}{dataset20} & \multicolumn{2}{c}{dataset100} \\
	   & 1C & 2C& 1C& 2C & 1C & 2C & 1C& 2C \\ 
		\hline
		ResNet18  & 16.1 & 14.1 & 20.5 & 20.2& 78.4 & 78.2 &  69.3 & 69.3\\
		ResNet34  & 17.4 & 16.6 & 21.2 & 20.6& 79.0 & 78.6 & 69.2 & 69.2\\
		ResNet50  & 16.6 & 15.2 & 20.7 & 20.7& 78.2 & 78.1 & 69.3 & 69.3\\
	 \bottomrule
 	\end{tabularx}
	\label{tb_ResultsR2scoreCNN}
\end{table}

\FloatBarrier

\end{document}